\documentclass[showpacs,aps,prb,reprint,superscriptaddress,nofootinbib,longbibliography]{revtex4-2}
\usepackage[colorlinks=true, pdfstartview=FitV,
bookmarks=true, bookmarksnumbered=true, breaklinks]{hyperref}
\usepackage[dvipdfmx]{graphicx}
\usepackage{amsmath,amssymb,bm,color,longtable,mathrsfs,slashed,comment,tikz}

\definecolor{darkviolet}{rgb}{0.58, 0.0, 0.83}
\definecolor{electricultramarine}{rgb}{0.25, 0.0, 1.0}
\definecolor{brightpink}{rgb}{1.0, 0.0, 0.5}
\hypersetup{linkcolor=brightpink, citecolor=electricultramarine, urlcolor=darkviolet}

\definecolor{lime}{HTML}{A6CE39}
\DeclareRobustCommand{\orcidicon}{
	\hspace{-3mm}
	\begin{tikzpicture}
	\draw[lime, fill=lime] (0,0) 
	circle [radius=0.16] 
	node[white] {{\fontfamily{qag}\selectfont \tiny ID}};
	\draw[white, fill=white] (-0.0625,0.095) 
	circle [radius=0.007];
	\end{tikzpicture}
	\hspace{-3mm}
}

\foreach \x in {A, ..., Z}{\expandafter\xdef\csname orcid\x\endcsname{\noexpand\href{https://orcid.org/\csname orcidauthor\x\endcsname}
			{\noexpand\orcidicon}}
}

\begin{document}
\begin{flushright}
\end{flushright}

\title{Randomized higher-order tensor renormalization group}

\author{Katsumasa~Nakayama\orcidA{}}
\email[]{katsumasa.nakayama@riken.jp}
\affiliation{RIKEN Center for Computational Science, Kobe 650-0047, Japan}

\date{\today}

\begin{abstract}
The higher-order tensor renormalization group (HOTRG) is a fundamental method to calculate the physical quantities by using a tensor network representation.
This method is based on the singular value decomposition (SVD) to take the contraction of all indices in the network with an approximation.
For the SVD, randomized singular value decomposition (R-SVD) is a powerful method to reduce computational costs of SVD.
However, HOTRG with the randomized method is not established.
%
%
We propose a randomized HOTRG method in a dimension $d$ with the computational cost $O(D^{3d})$ depending on the truncated bond dimension $D$.
We also introduce the minimally-decomposed TRG (MDTRG) as the R-HOTRG on the tensor of order $d+1$ with $O(D^{2d + 1})$ and a triad representation of the MDTRG (Triad-MDTRG) with $O(D^{d+3})$.
The results from these formulations are consistent with the HOTRG result with the same truncated bond dimension $D$.
\end{abstract}

\maketitle

\section{Introduction} \label{Sec:1}
The tensor renormalization group (TRG) calculation is a useful procedure to investigate physical systems on the lattice by representing a system as a tensor network, where the singular value decomposition (SVD) is used to approximately renormalize the system \cite{Levin:2006jai}.
The TRG can calculate the physical quantities, such as a partition function and its derivative, without the sign problem.
The TRG method produces high precision results, especially if the system has only sufficiently simple interaction.
Then we can study the details of the system, such as phase transitions and critical behaviors \cite{ Shimizu:2012wfa,Shimizu:2012zza,Yu:2013sbi,Zou:2014rha,Shimizu:2014uva,Shimizu:2014fsa,Unmuth-Yockey:2014afa,Takeda:2014vwa,Yang:2015rra,Kawauchi:2016xng,Sakai:2017jwp,Shimizu:2017onf,Kuramashi:2018mmi,Kadoh:2018tis,Kadoh:2018hqq,Bazavov:2019qih,Akiyama:2020ntf,Akiyama:2020soe,Akiyama:2021zhf,Akiyama:2021xxr,Bloch:2021mjw,CP1TRG,Akiyama:2021glo,Hirasawa:2021qvh,Bloch:2021rcq,Jha:2022pgy,Luo:2022eje,Akiyama:2022eip,Kuwahara:2022ubg,Akiyama:2023hvt}.

In a two-dimensional system, TRG has been improved to get more reliable results.
In order to reduce the computational cost and improve the precision which is estimated as $O(D^6)$ depending on the truncated bond dimension $D$, many approaches have been studied.
The anisotropic TRG (ATRG) \cite{ATRG} and Triad TRG (TTRG) \cite{TriadTRG} introduces additional decomposition to reduce the order of the cost to $O(D^{2d+1})$ and $O(D^{d+3})$ in $d$-dimensional system, respectively.
By using the different assumptions and approximation, we can calculate the contraction of the tensor network \cite{CTMRG, SRG, SRGtwo, TNR, LoopTNR, GILT,ProjectiveTRG,  O4CoreTRG,BTRG, GlobalTRG,TriadSRG, StocasticTRG,NuclearLoopTNR}.
The randomized TRG (R-TRG) is also an extension of the TRG, which uses the randomized SVD as a truncated SVD with the cost $O(D^5)$ \cite{RandTRG,rand_trunc}.

Some of these methods, including the original TRG, can not be applied in higher dimension. 
The higher-order TRG (HOTRG) is an approach to extend the TRG to higher dimensions \cite{HOTRG}.
HOTRG applies the SVD to find the isometry tensor which is a projector from an index range $D^2$ to $D$ as an approximated contraction.
We can find the isometry by the SVD and truncation of the small singular values.
The HOTRG in higher dimensions is used and produces reliable results in several systems.
On the other hand, since higher-dimensional systems have a larger degrees of freedom than two-dimensional systems, the computational cost is much more severe problem.
We can reduce the $O(D^{4d - 1})$ HOTRG to the $O(D^{2d - 1})$ ATRG and the $O(D^{d +3})$ TTRG, with the trade-off between the precision and computational cost.
These methods can produce precise results with reduced cost at large $D$.

Whereas the ATRG and TTRG apply the truncated SVD, the HOTRG does not, which is a disadvantage for the cost.
Because the further decompositions in ATGR and TTRG are also the systematic error resource and may be the origin of the non-monotonic convergence depending on $D$, the HOTRG without such a systematic error except for the truncation of the isometry will be useful.

In this paper, we introduce the truncated SVD to HOTRG which we call the {\it randomized HOTRG} (R-HOTRG).
The systematic error of the R-HOTRG is also dominated by the truncation of the isometry without other systematic errors.
We show that the R-HOTRG has the $O(D^{3d})$ cost and ${D^{d+1}}$ memory footprint except for the tensor $A$ of order $2d$.
In a three-dimensional system as an example of the higher dimension, we show the reduction of the calculation cost to $O(D^9)$.
The R-HOTRG is a fundamental approach similar to the R-TRG in the two dimension. 

We also introduce {\it minimally-decomposed TRG} (abbreviated as MDTRG hereafter) which is the R-HOTRG on the polyad tensor without any additional decompositions except for only once R-SVD for the contraction.
This MDTRG is done by $O(D^{2d + 1})$.
By introducing the several decompositions to the MDTRG, we can reduce the cost up to $O(D^{d+3})$.
The MDTRG reduces the cost without loss of precision compared to the original HOTRG.

We also introduce two general ideas for the general TRG method, which is also utilized for the MDTRG.
The first is the {\it isometry of the unit-cell tensor network} which is the sub-volume tensor network included in the approximation for the truncation.
The R-HOTRG and MDTRG take into account the contribution from the entire unit-cell tensor network, which helps us to improve the precision.
The second one is the {\it internal-line oversampling}.
This idea helps us to achieve the original HOTRG precision by using MDTRG.

\section{Randomized Higher order tensor renormalization group} \label{Sec:4}

We propose the R-HOTRG as the HOTRG with the randomized method.\footnote{The simple idea of the R-HOTRG is also mentioned in the appendix of \cite{TriadTRG}.}
The details of the HOTRG are in the paper \cite{HOTRG}, and also in the Appendix\ref{App:1}.

We consider a homogeneous three-dimensional tensor network constructed by the forth order tensor $A_{xyx'y'}$ as the square lattice tensor network, with the periodic boundary condition.
We also assume the elements of the tensor are real, without loss of generality.
For example, the partition function $Z$ is defined as
\begin{align}
&
Z\equiv\mathrm{Tr}\sum_{i} A_{x_iy_iz_ix_i 'y_i 'z_i '},
\end{align}
where $i$ is identified with the each lattice point and $\mathrm{Tr}$ represents the contraction of the all indices.

We find the coarse-grained tensor $A^\mathrm{(next)} _{XYzX'Y'z'}$ with the truncated indices $X,X',Y,$ and $Y'$ from the two-neighboring tensor $\Gamma_{x_1x_2x' _1 x' _2y_1y_2y' _1 y' _2 z_2z_1 '} ^{(AA)}=\sum_{\tilde{z}=1} ^DA_{x_1y_1\tilde{z}x' _1 y' _1 z' _1}A_{x_2y_2z_2x' _2 y' _2 \tilde{z}}$, where $D$ is the truncated bond dimension.
The $\tilde{z} = z_1 = z_2 '$ are summed from $1$ to $D$.
In this paper, we name $\Gamma$ the unit-cell tensor network.
The unit-cell tensor network $\Gamma ^{(AA)}$ becomes the coarse-grained tensor $A^\mathrm{(next)}$ at one step.
The partition function of the system is represented by the $A^\mathrm{(next)}$ as $Z\simeq\mathrm{Tr}\sum_{i} A^\mathrm{(next)} _{X_iY_iz_iX_i 'Y_i 'z_i '}$ in a half volume.

As the coarse-graining of $x$-direction, we have to find the tensor $U_{x_1x_2X} ^{(\vec{x})}$ with the index $X = 1,...,D$, which is called the isometry.
The isometry approximates the $D^2$ contraction of the $x_1$ and $x_2$ by $X$ with the bond dimension $D$.
In order to find the approximated isometry, we use the SVD of $\Gamma\Gamma^t$.
\begin{align}
&
\sum_{x' _1, x' _2, y,z_2,z' _1 = 1}^D\Gamma_{[x_1x_2][x' _1 x' _2 ,y,z_2z_1 ']} ^{(AA)}\Gamma^{(AA)} _{[x_1 ^tx_2 ^t][x' _1 x' _2 ,y,z_2z_1 ']} \nonumber\\
&
=\sum_{k=1} ^{D^2}U_{x_1x_2 k} ^{(\vec{x})}\lambda_kU ^{(\vec{x})} _{x _1 ^tx _2 ^t k},
\label{iso1}
\end{align}
with the singular value $\lambda_k$,
where $\sum_{y}$ is an abbreviation of $\sum_{y _1, y _2,y' _1, y' _2}$.
The tensor $\Gamma\Gamma^t$ can be written by the tensor $A$ as followings.
\begin{align}
&
\sum_{x' _1, x' _2, y,z_2,z' _1 = 1}^D\Gamma_{[x_1x_2][x' _1 x' _2 ,y,z_2z_1 ']} ^{(AA)}\Gamma^{(AA)} _{[x_1 ^tx_2 ^t][x' _1 x' _2 ,y,z_2z_1 ']} \nonumber\\
&
=
\sum_{x' _1, x' _2, y,z_2,z' _1 = 1}^D
\sum_{\tilde{z},\tilde{z} ^t = 1}^D
A_{x_1y_1\tilde{z}x' _1 y' _1 z' _1}
A_{x_2y_2z_2x' _2 y' _2 \tilde{z}}\nonumber\\
&
\ \ \ \ \ \ \ \ \ \ \ \ \ \ \ \ \ \ \ \ \ \ \ \ \ 
\times
A_{x_1 ^ty_1\tilde{z}^tx' _1 y' _1 z' _1} 
A_{x_2 ^t y_2z_2x' _2 y' _2 \tilde{z}^t} 
\label{iso2}
\end{align}

In this paper, we sometimes introduce the square bracket to guide the contracted indices and matrices.
We also calculate the isometry of the $y$-direction $U ^{(\vec{y})}$ by the SVD of $[\Gamma\Gamma^t]_{y_1y_2y_1 ^ty_2 ^t}$.
By using the isometry $U ^{(\vec{x})}$ and $U ^{(\vec{y})}$, we take the contraction of the tensor,
\begin{align}
A^{(\mathrm{next})} _{XYzX'Y'z'}=\sum_{x,y=1} ^D 
&U_{x_1 'x_2 'y_1 'y_2 'X'Y'} 
U_{x_1x_2y_1y_2XY} \nonumber\\
&\times
\Gamma_{x_1x_2x' _1 x' _2y_1y_2y' _1 y' _2 z_2z_1 '} ^{(AA)}.
\label{cont1}
\end{align}
where $\sum_{x}$ is an abbreviation of $\sum_{x _1, x _2,x' _1, x' _2}$.
We introduce the product of the $U ^{(\vec{x})} _{x_1x_2X}$ and $U_{y_1y_2Y} ^{(\vec{y})}$ as $U_{x_1x_2y_1y_2XY} = U ^{(\vec{x})} _{x_1x_2X}U_{y_1y_2Y} ^{(\vec{y})}$ .
The computational cost of the isometry step (\ref{iso1})-(\ref{iso2}) and contraction step (\ref{cont1}) are $O(D^8)$ and $O(D^{11})$, respectively.
In the contraction step, the bottleneck is the contraction of the $U ^{t}A$ and $AU$, 

\begin{align}
&
[U ^{t}A]_{[XYz_2][x_1x_2 'y_1y_2 '\tilde{z}]} = \sum_{x _2,y _2 = 1} ^DU_{x_1x_2y_1y_2XY}  A_{x_2y_2z_2x' _2 y' _2 \tilde{z}} ,\\
&
[A U^{}]_{[x_1x_2 'y_1y_2 '\tilde{z}][X'Y'z_1 ']} = \sum_{x _1' ,y_1 '  = 1} ^DA_{x_1y_1\tilde{z}x' _1 y' _1 z' _1} U_{x_1 'x_2 'y_1 'y_2 'X'Y'},\\
&
A^{(\mathrm{next})} _{XYzX'Y'z'}\nonumber\\
&=
\sum_{x_1, x_2 ' ,y_1, y_2 ' , \tilde{z} = 1} ^D
[U ^{t}A]_{[XYz_2][x_1x_2 'y_1y_2 '\tilde{z}]}
[A U^{}]_{[x_1x_2 'y_1y_2 '\tilde{z}][X'Y'z' _1]},
\label{simple_cont}
\end{align}
where the cost is $O(D^{11})$.

In the contraction step, we introduce the randomized method to reduce the cost to $O(D^{9})$.
The R-HOTRG use the following contraction instead of (\ref{simple_cont}).
\begin{align}
&
A^{(\mathrm{next})} _{XYzX'Y'z'}\nonumber\\
&=
\sum_{x,y, z_2, \tilde{z},\tilde{X},\tilde{Y} = 1} ^D
\sum_{\omega = 1} ^{rD}
Q_{XYz\omega} Q_{\tilde{X}\tilde{Y}z_2\omega} ^*.
\nonumber\\
&
\ \ \ \ \ \ \ \ \ \ \ \ \ \ \ \ \ \ \ \ \times
U  _{x' _1x' _2 y' _1y' _2 X'Y'} 
U  _{x_1x_2 y_1y_2 \tilde{X}\tilde{Y}} 
\nonumber\\
&
\ \ \ \ \ \ \ \ \ \ \ \ \ \ \ \ \ \ \ \ \times
A_{x_1y_1\tilde{z}x' _1 y' _1 z' _1}
A_{x_2y_2z_2x' _2 y' _2 \tilde{z}}
\label{R-HOTRG}
\end{align}
where we introduced the orthogonal matrix $Q$ from the R-SVD.
In order to calculate $Q$, we consider the random matrix $\Omega_{X' Y' z'\omega}$ with the $\omega = 1,...,rD$  with the oversampling constant $r \geq 1$.
We take the sequential contraction as follows to calculate the sample matrix $\Theta$.
\begin{align}
&
[U \Omega ]_{x' _1 y' _1 z_1 ' x' _2 y' _2 \omega }=
\sum_{X' , Y' = 1} ^D
U_{[x' _1 x' _2  y' _1 y' _2] [X'Y']}
\Omega_{[X' Y'] [z_1 ' \omega]},\\
&
[AU \Omega]_{x' _2y' _2\tilde{z} x_1  y_1  \omega} \nonumber\\
&=
\sum_{x' _1,y' _1,z_1' = 1} ^D
A_{[x_1y_1\tilde{z}][x' _1y' _1z_1 ']}
[U \Omega ]_{[x' _1y' _1z_1 '] [x' _2y' _2 \omega] },\\
&
[AAU \Omega]_{x _1 x _2y _1 y _2 z_2 \omega }\nonumber\\
&=
\sum_{x' _2,y' _2, \tilde{z} = 1} ^D
A_{[x_2y_2z_2][x' _2 y' _2 \tilde{z}]}
[AU \Omega]_{[x' _2y' _2\tilde{z}] [x_1  y_1  \omega]},\\
&
\Theta_{XY z_2 \omega}\nonumber\\
&=
\sum_{x_1, x _2,y_1, y _2, = 1} ^D
U _{[x _1 x _2  y _1 y _2] [XY]}
[AAU \Omega]_{[x _1 x _2y _1 y _2] [z_2 \omega]}
\end{align}
After we obtain the tensor $\Theta$, we calculate the $Q$ from tensor $\Theta$ by using QR decomposition, $\Theta_{[XYz] \omega} =\sum_{m = 1} ^{rD} Q_{XYzm} R_{m\omega}$. This orthogonal matrix $Q$ is used in R-HOTRG as shown in Eq. (\ref{R-HOTRG}).

The backward contraction of $Q$ is as follows.
\begin{align}
&
[Q^\dagger U ^{t}]_{ x _1y _1\omega x _2 y _2z_2}=
\sum_{\tilde{X},\tilde{Y} = 1} ^D
Q_{[\tilde{X}\tilde{Y}][z_2\omega]} ^*
U_{[x _1 x _2y _1 y _2 ][\tilde{X}\tilde{Y}]} ,\\
&
[Q^\dagger U ^t A]_{x' _2 y' _2\omega  x _1y _1\tilde{z}}\nonumber\\
&
=
\sum_{x_2,y_2,z_2 = 1} ^D
[Q^\dagger U ^{t}]_{ [x _1y _1\omega] [x _2 y _2z_2]}
A_{[x_2y_2z_2][x' _2 y' _2 \tilde{z}]},\\
&
[Q^\dagger U ^tAA]_{ z_1 ' \omega, x' _1x' _2y' _1y' _2}\nonumber\\
&
=
\sum_{x_1,y_1,\tilde{z}  = 1} ^D
[Q^\dagger U ^t A]_{[x' _2 y' _2\omega]  [x _1y _1\tilde{z}]}
A_{[x_1y_1\tilde{z}][x' _1y' _1z_1 ']},\\
&
\Lambda _{X'Y' z\omega }\nonumber \\
&
=
\sum_{x_1 ' ,x_2 ' ,y_1 ' ,y_2 ' = 1} ^D
[Q^\dagger U ^tAA]_{[z_1 ' \omega][x' _1x' _2y' _1y' _2]}
U _{[x_1 ' x_2 'y_1 ' y_2 '] X'Y'},\\
&
A^{(\mathrm{next})} _{XYz_2X'Y'z_1 '}
=
\sum_{\omega = 1} ^{rD}
Q_{XYz_2\omega}
\Lambda _{X'Y' z_1 '\omega }.
\end{align}
We define the tensor $\Lambda$.
This procedure is based on the randomized SVD (R-SVD) method \cite{RandTRG,TriadTRG}.
More details of the randomized SVD are also discussed in \cite{RandTRG,TriadTRG}, and a theoretical discussion is shown in \cite{rand_trunc}.
We also briefly summarize the R-SVD procedure in Appendix \ref{App:2}.

Since the maximum order tensor in the total procedure is still sixth-order which is the same as the original tensor $A$, the memory footprint is also still $O(D^6)$, naively.
In addition, we can reduce the memory footprint by using loop-blocking technique as similar manner of HOTRG. 
We store the intermediate tensors $\Theta $ and $Q$ of order $d+1$, which requires the $O(D^4)$ memory usage.

\begin{figure}[tb!]
 \centering
 \includegraphics[clip,width=1.0\columnwidth]{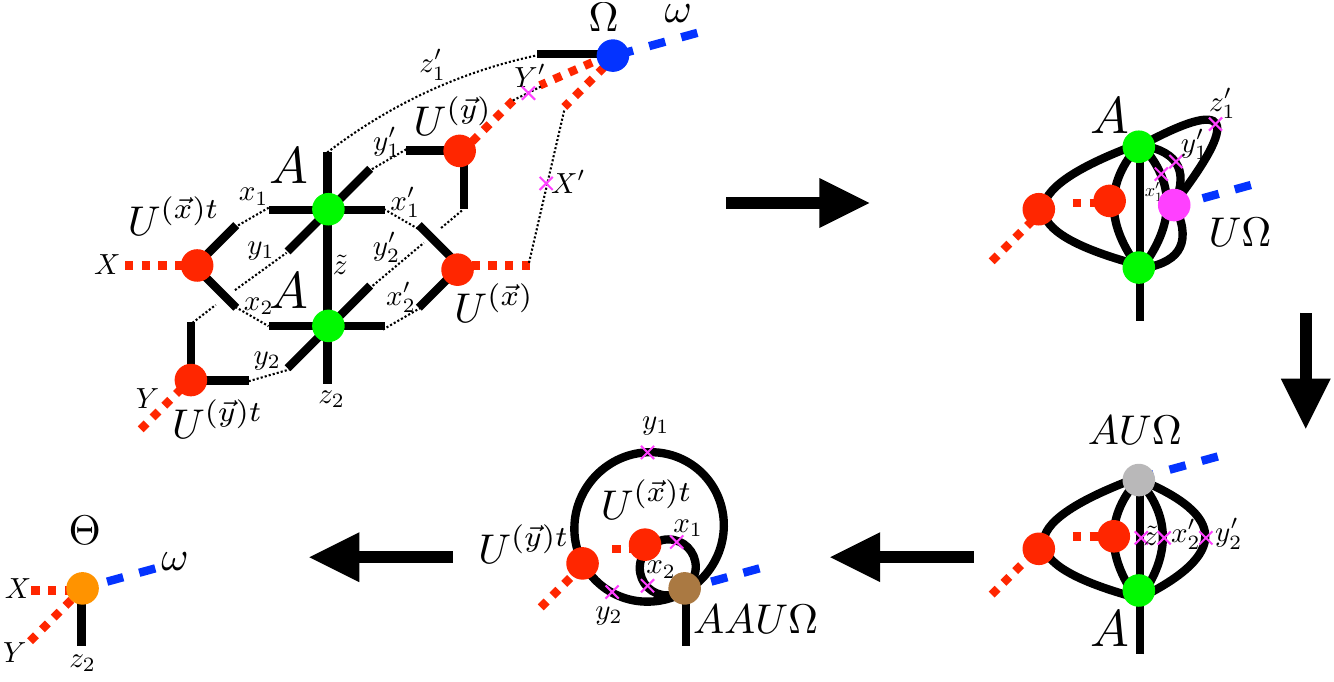}
 \caption{Schematic picture of the $\Theta = U^tAAU\Omega$ calculation of the three-dimensional R-HOTRG. The index with a cross point is contracted to the next step.}
 \label{disp}
\end{figure}
\begin{figure}[tb!]
 \centering
 \includegraphics[clip,width=1.0\columnwidth]{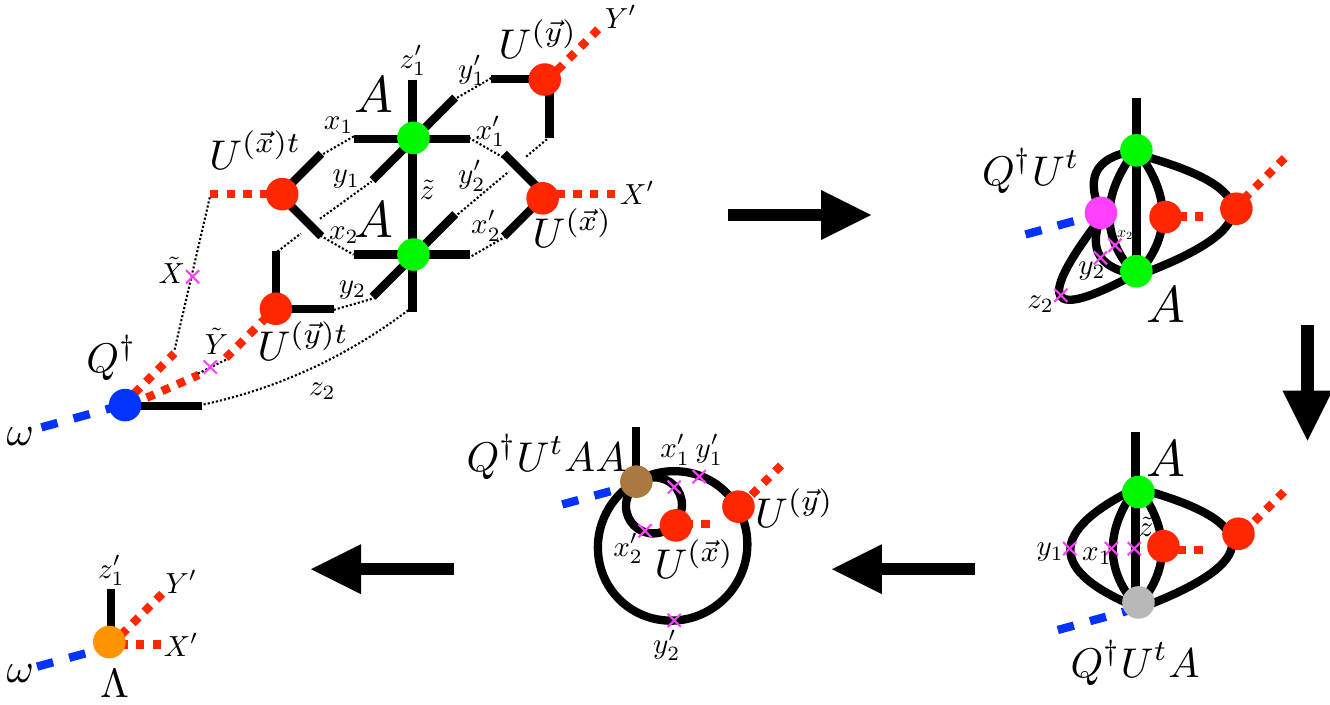}
 \caption{Schematic picture of the $\Lambda = Q^\dagger U^tAAU$ calculation of the three-dimensional R-HOTRG. The index with a cross point is contracted to the next step.}
 \label{disp2}
\end{figure}

We can estimate the cost of the contractions for each step.
The dominant part of the cost is the contraction of the $[AU\Omega]$, $[AAU\Omega]$, $[Q^\dagger U^tA]$, and $[Q^\dagger U^tAA]$. 
This part requires the $O(D^9)$ cost, and then the R-HOTRG in the 3-dimensional system requires the $O(D^{9})$ cost.
We show the schematic picture of the key steps to calculate tensor $\Theta$ and $\Lambda$ in three-dimensional R-HOTRG in Figs. \ref{disp} and $\ref{disp2}$, respectively.

In a general $d$-dimensional system, the original tensor $A$ of order $2d$ is stored in the calculation. 
The dominant part of the cost is still the contraction of the $[AU\Omega]$, $[AAU\Omega]$, $[Q^\dagger U^tA]$, and $[Q^\dagger U^tAA]$. 
This part requires the $O(D^{3d})$ cost.
The isometry can be calculated with the cost $O(D^{2d+2})$ with no truncated SVD, which is not the dominant part of the calculation time.

We mention the pre-factor of the computational cost that comes from the randomized SVD \cite{RandTRG,rand_trunc} in the R-HOTRG.
Since the R-HOTRG includes contraction with an oversampled index in $O(D^{3d})$ contraction, the cost including the pre-factor becomes $O(D^{4d - 1}) \rightarrow O(rD^{3d})$.
In addition to this oversampling parameter, we can introduce the power iteration scheme to improve the approximation \cite{rand_trunc,RandTRG}.
If we calculate QR decomposition $q$ times in this scheme, the pre-factor becomes $O(qrD^{3d})$.

Note that we can control the cost by changing the truncated bond dimension at the randomized matrix product step.
For example, in a three-dimensional system, if we truncate the bond dimension up to $D^2$ (not $D$), the computational cost becomes $O(D^{10})$ which is still better than the original HOTRG.
In general, we can choose the cost in the range of $O(D^{3d})$ and original $O(D^{4d - 1})$ by changing the truncated bond dimension for the randomized method.
Thus we can choose the trade-off between the precision and cost.

\section{Minimally-decomposed TRG} \label{Sec:4p1}
We consider the cost reduction of the R-HOTRG.
In the R-HOTRG procedure, $\Lambda$ and $Q$ are the tensor of order $d+1$ which constructs the next tensor $A^\mathrm{(next)} = Q\Lambda ^t$.
For the memory usage and the computational cost, tensors of lower order is preferable.
In this section, we do not calculate the $A^\mathrm{(next)}$, and redefine the $\Gamma$ by the SVD of $\Lambda$.

Let us define the tensors $E,F,G$ and $H$ of order $d+1$ defined by the SVD of the $\Lambda_{x'y'z'\omega} = \sum_{k=1} ^{rD}\overline{U}_{x'y'z' k}{s}  _k\overline{V}_{\omega k}$ as $E = \overline{U}$, $F = Q\overline{V}{s}$, $G = \overline{U}{s}$ and $H = Q\overline{V}$.
We consider these tensors $E,F,G$ and $H$ construct the fundamental unit-cell tensor network $\Gamma$ as follows, 
\begin{align}
&
\Gamma ^{(EFGH)}= 
\sum_{\tilde{z}=1} ^D
\sum_{e,g= 1} ^{rD}E_{x_1 'y_1 'z_1 ' e}F_{x_1 y_1 \tilde{z}e}G_{x_2 'y_2 '\tilde{z} g}H_{x_2 y_2 z_2 g},
\end{align}
which means the replacement $\Gamma^{(AA)} = AA \rightarrow \Gamma^{(EFGH)} = EFGH$.
Hereafter, we call this R-HOTRG with the unit-cell tensor network $\Gamma^{(EFGH)}$ the minimally decomposed TRG (MDTRG), since the decomposition for each step is taken only once.
Note that the oversampled lines $e$ and $g$ remain larger $rD$ than $D$.
We call this method the internal-line oversampling to distinguish it from the conventional oversampling for the RSVD. 
We will see how this internal-line oversampling works in the Sec.\ref{Sec:5}.

\begin{figure}[tb!]
 \centering
 \includegraphics[clip,width=1.0\columnwidth]{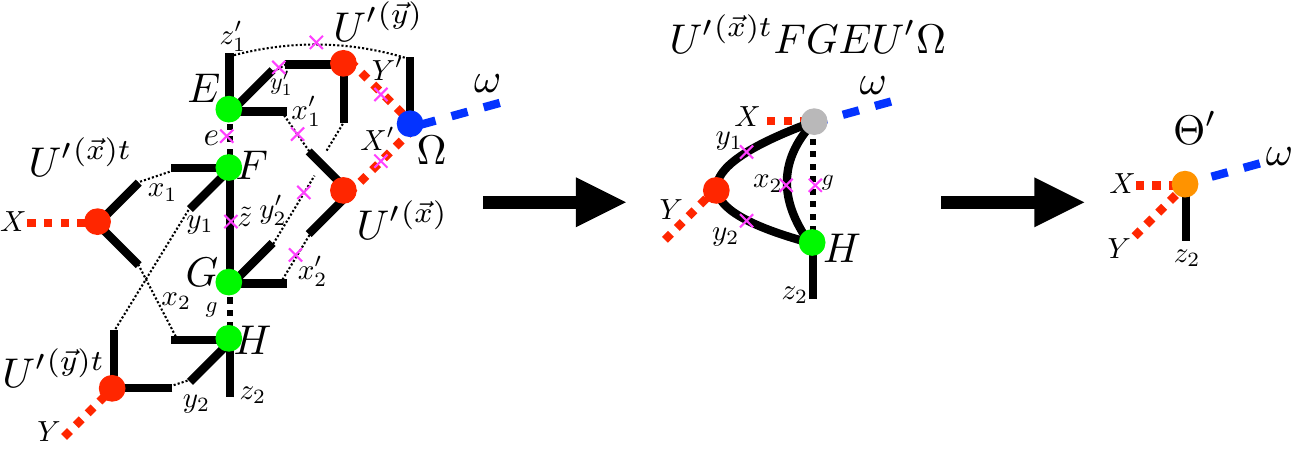}
 \caption{Schematic picture of the contraction of the $\Theta'=U^{(y)}HU^{(x)}FGEU^{t}\Omega$ calculation of the MDTRG in a three-dimensional system. The index with a cross point is contracted to the next step.}
 \label{3d_tetra}
\end{figure}

\begin{figure}[tb!]
 \centering
 \includegraphics[clip,width=1.0\columnwidth]{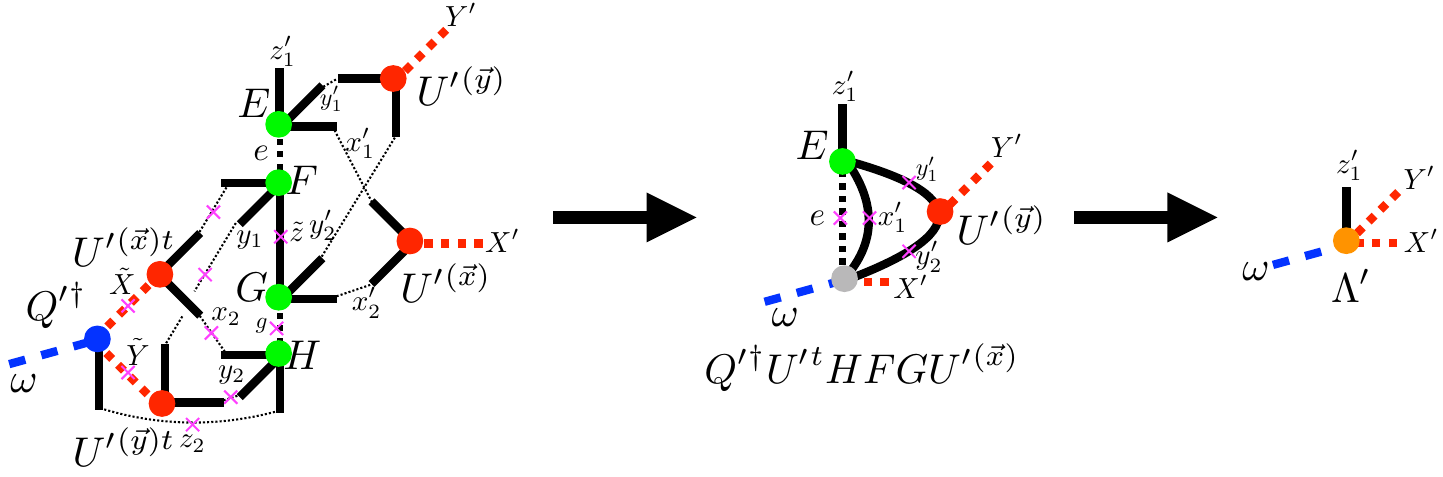}
 \caption{Schematic picture of the contraction of the $\Lambda'=Q^\dagger U' {}^t HFGU' {}^{(x)}EU' {}^{(y)}$ calculation of the MDTRG in a three-dimensional system. The index with a cross point is contracted to the next step.}
 \label{3d_tetra2}
\end{figure}

After this replacement, the isometry step is almost the same by using $\Gamma^{(EFGH)}\Gamma^{(EFGH)t}$.
We define the isometry $U'$ by using the SVD of the $\Gamma^{(EFGH)}\Gamma^{(EFGH)t}$.
The contraction step is also not drastically changed.
Since the fundamental unit-cell tensor is now $\Gamma^{(EFGH)} = EFGH$, we do not calculate the tensor $A^{\mathrm{(next)}}$ of order $2d$.
We consider the randomized SVD of $\Gamma^{(EFGH)}$ as follows, 
\begin{align}
&\sum_{\omega = 1} ^{rD}\overline{U}_{X'Y'z_1 ' \omega} ^{(EFGH)} \overline{s} _\omega\overline{V} ^{(EFGH)} _{XYz_2\omega}
=\nonumber \\
\sum_{x,y,\tilde{z} = 1} ^{D}
\sum_{e,g = 1} ^{rD}
&
U_{x_1x_2 y_1y_2XY} '
U_{x_1 'x_2 'y_1 'y_2 'X'Y'} ' \nonumber\\
\times 
&E_{x_1 ' y_1 ' z_1 ' e} F_{x_1y_1\tilde{z}e} G_{x_2 ' y_2 ' \tilde{z} g} H_{x_2y_2z_2 g}. 
\end{align}
The $\sum_{x,y}$ is an abbreviation of $\sum_{x_1, x_2, x_1 ', x_2 ' ,y_1, y_2, y_1 ', y_2 ' }$.
We define the next tensor as $E^{(\mathrm{next})} = \overline{U} ^{(EFGH)}$, $F^{(\mathrm{next})} =( \overline{V} ^{(EFGH)}\overline{s})$, $G^{(\mathrm{next})} = (\overline{U} ^{(EFGH)}\overline{s})$ and $H^{(\mathrm{next})} = \overline{V} ^{(EFGH)}$.
We show the schematic picture of the MDTRG in three dimensions in Figs. \ref{3d_tetra} and \ref{3d_tetra2}.


In this method, the isometry step needs $O(r^2D^{d + 3})$ computational cost, and contraction step needs $O(r^2D^{2d + 1})$.
Note that if we do not introduce the internal-line oversampling, the pre-factor $r^2$ becomes $r$.
The calculation has to store the tensors $E, F, G,$ and $H$ which are tensors of order $d+1$.
We store the intermediate tensors $\Theta' $ and $Q'$ of order $d+1$, which requires the $O(D^{d+1})$ memory footprint.

We mention the initial preparation of the tensors $E,F,G,$ and $H$ of order $d+1$ for the unit-cell tensor network $\Gamma^{(EFGH)}$.
We can find this representation from the canonical polyadic representation which is also utilized in TTRG \cite{TriadTRG}.
In addition, if the initial tensor network has a sufficiently small index, we can find it by the full SVD of $A$ with no any truncations.
We also introduce the method for the initial preparation with respect to the unit-cell tensor by the discussion of the approximation which takes into account the contribution originated from the unit-cell tensor $\Gamma$ in Appendix\ref{App:3}.


\section{MDTRG with Triad representation} \label{Sec:4p7}
In this section we reduce the computational cost from $O(D^{2d + 1})$ by additional decompositions.
In order to take into account the contribution of the unit-cell tensor network $\Gamma^{(EFGH)} = EFGH$ to the MDTRG, we carefully introduce the isometry for this additional decomposition steps.

Before the isometry and contraction step, we decompose the tensors $E, F,G,$ and $H$ of order $d+1$ to the third order tensors, called triad tensors in TTRG \cite{TriadTRG}.
For simplicity, we introduce the procedure to get the triad representation from the tensor of order $d+1$ in a three-dimensional system with the $z$-direction contraction $\sum_{\tilde{z}=1} ^DF_{xy\tilde{z}e}G_{x'y'\tilde{z}g}$.

\begin{figure}[tb!]
 \centering
 \includegraphics[clip,width=1.0\columnwidth]{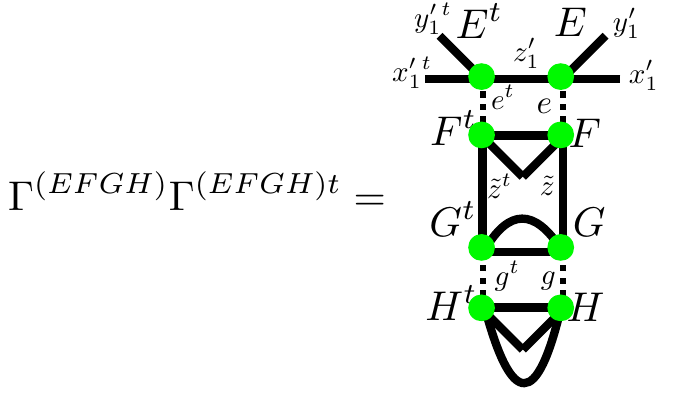}
 \caption{Schematic picture of the calculation of the isometry $U_{x'y'k} ^{(\vec{i})}$ defined by the SVD of $\Gamma^{(EFGH)}\Gamma^{(EFGH)t} = U ^{(\vec{i})}\lambda^{(\vec{i})}U ^{(\vec{i})t}$ for the Triad-MDTRG in a three-dimensional system. }
 \label{3d_triad_initial}
\end{figure}

We consider the isometry $U_{x'y'k} ^{(\vec{i})}$ defined by the corresponding SVD of $\Gamma^{(EFGH)}\Gamma^{(EFGH)t} = U ^{(\vec{i})}\lambda^{(\vec{i})}U ^{(\vec{i})t}$,
\begin{align}
&
\sum_{x_1,x_2,  x_2 ' ,y_1 ,y_2 , y_2 ' ,z_2 ,z_1 ' = 1} ^D
\Gamma_{[x_1 'y_1 '] [x_1x_2  x_2 ' y_1 y_2  y_2 ' z_2 z_1 ']} ^{(EFGH)}\nonumber\\
&
\ \ \ \ \ \ \ \ \ \ \ \ \ \ \ \times
\Gamma_{[x_1x_2  x_2 ' y_1 y_2  y_2 ' z_2 z_1 '] [x_1 ' {}^ty_1 '{}^t]} ^{(EFGH)} \nonumber\\
&
\ \ \ \ \ \ \ \ \ \  \ \ \ \ \ =
\sum_{i = 1} ^{rD}
 U^{(\vec{i})} _{x_1 ' y_1 '  i} \lambda^{(\vec{i})} _{i}U^{(\vec{i})} _{x_1 ' {}^ty_1 ' {}^t i},
 \end{align}
and we define the triad tensor $I =U^{(\vec{i})} $ and $J_{z' i j} = \sum_{x', y' = 1} ^D U^{(\vec{i})} _{x' y' i}E_{x' y' z' j} $.
Figure \ref{3d_triad_initial} shows the $\Gamma^{(EFGH)}\Gamma^{(EFGH)t}$ for the $U ^{(\vec{i})}$.
To get the triad representation, we perform the same procedure to the other $F,G,$ and $H$ and define corresponding isometries $U ^{(\vec{k})}, U ^{(\vec{m})}, $ and $U ^{(\vec{o})}$, respectively.
We define the triad tensors $K,L,M,N,O,P$, and summarize it with $I$ and $J$ as follows, 
\begin{align}
&
I =U^{(\vec{i})} ,\nonumber\\
&
J_{z' i j} = \sum_{x', y' = 1} ^D U^{(\vec{i})} _{x' y' i}E_{x' y' z' j} ,\nonumber\\
&
K_{z j k} = \sum_{x, y = 1} ^D U^{(\vec{k})} _{x y k}G_{x y z j} ,\nonumber\\
&
L =U^{(\vec{k})} , \nonumber\\
&
M =U^{(\vec{m})} ,\nonumber\\
&
N_{z' m n} = \sum_{x', y' = 1} ^D U^{(\vec{m})} _{x' y' m}E_{x' y' z' n}, \nonumber\\
&
O_{z m o} = \sum_{x, y = 1} ^D U^{(\vec{o})} _{x y o}G_{x y z m} ,\nonumber\\
&
P =U^{(\vec{o})}.
 \end{align}
The computational cost is $O(D^{d+3})$, which is the same order as the TTRG.
After introducing the triad representation, we calculate the isometry $U''  = U'' {}^{(\vec{x})} U'' {}^{(\vec{y})}$ in the same manner as the TTRG by the SVD of $IJKLMNOP$. The cost is $O(D^6)$ without randomized SVD.

\begin{figure}[tb!]
 \centering
 \includegraphics[clip,width=1.0\columnwidth]{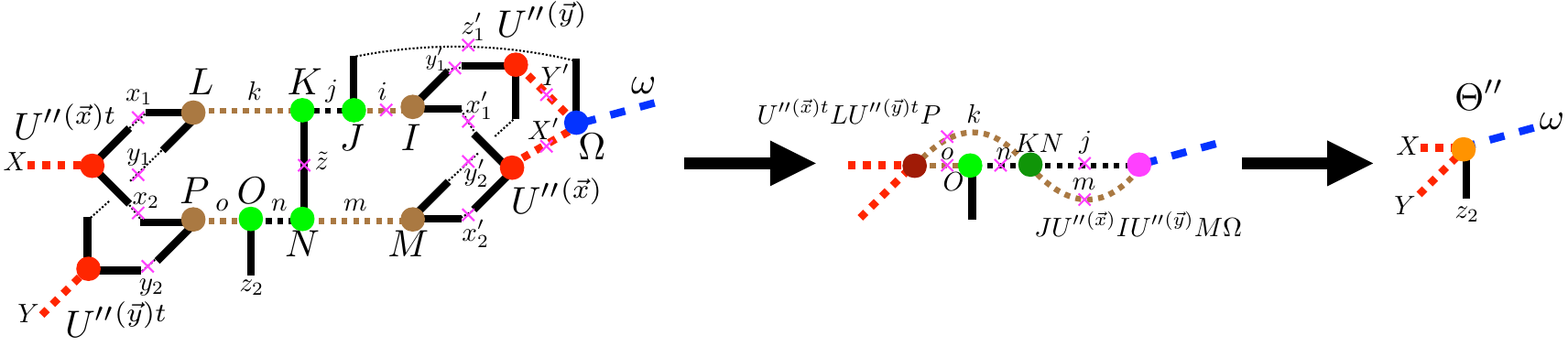}
 \caption{Schematic picture of the contraction of the $\Theta''$ of the Triad-MDTRG in a three-dimensional system.The index with a cross point is contracted to the next step.}
 \label{3d_triad}
\end{figure}
\begin{figure}[tb!]
 \centering
 \includegraphics[clip,width=1.0\columnwidth]{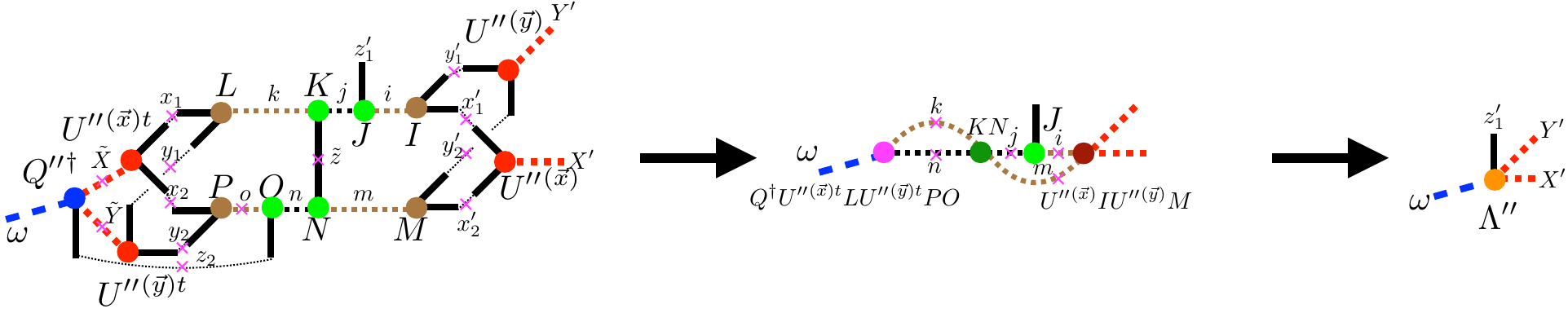}
 \caption{Schematic picture of the contraction of the $\Lambda''$ of the Triad-MDTRG in a three-dimensional system. The index with a cross point is contracted to the next step.}
 \label{3d_triad2}
\end{figure}

In the contraction step, we consider the randomized method to get the next tensors $E^\mathrm{(next)}, F^\mathrm{(next)}, G^\mathrm{(next)}$, and $H^\mathrm{(next)}$ of order $d+1$.
Figures \ref{3d_triad} and \ref{3d_triad2} shows the schematic pictures of the contraction steps of the MDTRG with the triad representation.
We consider the following contraction with the isometry $U^{(IJKLMNOP)}$,
\begin{align}
&\sum_{\omega = 1} ^{rD}
\overline{U}^{(IJKLMNOP)} _{X'Y'z' \omega}\overline{s}' _{\omega}\overline{V} ^{(IJKLMNOP)} _{XYz\omega}
\nonumber \\
&=\sum_{\{x,y,\tilde{z} = 1\}} ^D
\sum_{\{i,j,k,m,n,o = 1\}} ^{rD}
U''   _{x_1x_2 y_1y_2XY} 
U  ''  _{x_1 'x_2 'y_1 'y_2 'XY}  \nonumber\\
&
\ \ \ \ \ \ \ \ \ \ \ \ \ \ \ \ \ \ \ \ \ \ \ \ \ \ \ \ \times 
I_{x_1 ' y_1 ' i} J_{z' _1 i j} K_{\tilde{z} j k} L_{x_1 y_1 k}\nonumber\\
&
\ \ \ \ \ \ \ \ \ \ \ \ \ \ \ \ \ \ \ \ \ \ \ \ \ \ \ \ \times 
M_{x_2 ' y_2 ' m} N_{\tilde{z} m n} O_{z_2 n o} P_{x_2 y_2 o}
\end{align}
In the MDTRG with the triad representation (abbreviated as Triad-MDTRG), we define the next tensor as $E^{(\mathrm{next})} = \overline{U} ^{(IJKLMNOP)}$, $F^{(\mathrm{next})} =( \overline{V} ^{(IJKLMNOP)}\overline{s}')$, $G^{(\mathrm{next})} = (\overline{U} ^{(IJKLMNOP)}\overline{s}')$ and $H^{(\mathrm{next})} = \overline{V}^{(IJKLMNOP)} $.
The Triad-MDTRG still takes into account the whole unit-cell tensor $\Gamma^{(EFGH)}$ in the contraction step, by using the randomized SVD.

This method has many similarities to the original TTRG.
We note the typical properties of the Triad-MDTRG as follows.

\begin{itemize}
\item One time R-SVD---Triad-MDTRG use R-SVD only one time for each renormalization step.
\item Respecting the unit-cell tensor network--- Each procedure in the Triad-MDTRG approximates the whole part of the unit-cell tensor $\Gamma^{(EFGH)}$.
Our calculation will be converged to the result of the HOTRG.
\item The unit-cell tensor network is defined by the tensor of order $d+1$---The unit-cell tensor network of MDTRG is the $\Gamma^{(EFGH)}$ which is constracted by the tensors $E, F, G$, and $H$ of order $d+1$, not the triad order. This setup help us to respect the unit-cell tensor network $\Gamma^{(EFGH)}$.
\item The internal-line oversampling---In order to achieve the convergence to the result of the HOTRG, we take the oversampling of the internal lines $\{i,j,k,m,n,o\}$.
\item $O(r^3D^6)$ computational cost---The Triad-MDTRG needs $O(r^3D^6)$ computational cost in the three-dimension with the internal-line oversampling.
\end{itemize}

Table \ref{tab:comp} summarizes the numerical cost and the order of the tensor constructing the unit-cell tensor $\Gamma$ for several methods.

\begin{table}[t!]
 \centering
 \caption{Comparison of computational cost and order of the tensor in the unit-cell tensor network $\Gamma$ in TRG.}
\begin{tabular}{l l l } 
 \hline \hline
   & Cost & Order \\ 
\hline
2d-TRG \cite{Levin:2006jai}   &$O(D^6)$ & $4\sim 3$ \\
2d-R-TRG \cite{RandTRG}  &$O(D^5)$ & $4\sim 3$ \\
HOTRG \cite{HOTRG}& $O(D^{4d-1})$ & $2d$ \\
R-HOTRG(This work) & $O(D^{3d})$ & $2d$ \\
ATRG \cite{ATRG}& $O(D^{2d-1})$ & $2d\sim d+1$ \\
MDTRG(This work) & $O(D^{2d-1})$ & $d+1$ \\
TTRG \cite{TriadTRG} & $O(D^{d+3})$ & $3$ \\
Triad-MDTRG(This work) & $O(D^{d+3})$ & $d+1$ \\
\hline \hline  
 \end{tabular}
 \label{tab:comp}
\end{table}

\section{Numerical calculation in the Ising model} \label{Sec:5}

We apply the R-HOTRG method to the three-dimensional Ising model.
The partition function $Z$ is defined as
\begin{align}
&
Z\equiv\mathrm{Tr}\sum_{i} A_{x_iy_iz_ix_i 'y_i 'z_i '},
\end{align}
where $i$ is identified with the each lattice point and $\mathrm{Tr}$ represents the contraction of the all indices.
We also introduce the tensor $A$ for the Ising model, 
\begin{align}
&
A_{xyzx'y'z'}
=
\sum_{k = 1} ^2
W_{kx}
W_{ky}
W_{kz}
W_{kx'}
W_{ky'}
W_{kz'},
\end{align}
where the $W$ is a $2\times 2$ matrix with inverse temperature $\beta$, 
\begin{align}
&
W_{kx} = 
\begin{pmatrix}
\sqrt{\mathrm{cosh}\beta} & \sqrt{\mathrm{sinh}\beta} \\
\sqrt{\mathrm{cosh}\beta} & -\sqrt{\mathrm{sinh}\beta}\\ 
\end{pmatrix}.
\end{align}
For MDTRG and Triad-MDTRG, we prepare frist $\Gamma^{(EFGH)}$ by R-SVD as introduced in Appendix\ref{App:3}.
The volume size is calculated up to $V = 2^{45}$ at critical point $T_c = 4.5115 = \beta^{-1} _c$ and we compute the free energy density $\mathcal{F}\equiv - \frac{1}{\beta V}\mathrm{log}Z$. 
We also take the rotation of the indices $\{x,y,z\} \rightarrow \{y,z,x\}$ after each renormalization steps, $\Gamma\rightarrow \Gamma^\mathrm{(next)}$.

In the randomized matrix product step, we introduced the oversampling parameter $r=6$ for $D^3\times rD$ random matrix $\Omega$, with the QR decompositions $q=2$ times in randomized SVD.
Our calculations are performed on the Apple M2 and use the library Eigen for matrix product and decomposition.

\begin{figure}[t]
 \centering
 \includegraphics[clip,width=1.0\columnwidth]{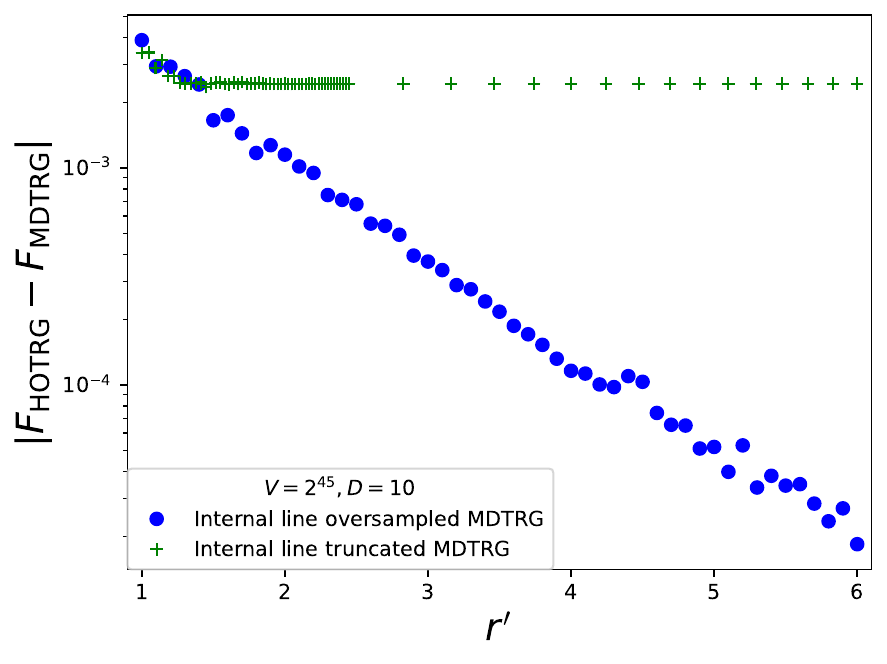}
 \caption{The difference of the free energy from the HOTRG in the three-dimensional Ising model at a critical temperature depending on the typical oversampling parameter $r'$ by the MDTRG. The typical oversampling parameter $r'$ is $r' = r$ for the MDTRG (dot) and $r' = \sqrt{r}$ for the MDTRG with no internal-line oversampling (plus).}
 \label{R_dep}
\end{figure}

\subsection{Internal-line oversampling}
First, we show the effectiveness of the internal-line oversampling in Fig. \ref{R_dep} by the MDTRG with the parameters, $q=2$, and $D = 10$.
Since the MDTRG requires the $O(r^2D^7)$ and original (no internal-line oversampled) MDTRG requires the $O(rD^7)$ computational cost, we define the typical oversampling parameter $r'$ for the randomized method.
The typical oversampling parameter $r'$ of the MDTRG is $r' = r$, and that of no internal-line oversampled MDTRG is $r' = \sqrt{r}$.
Figure \ref{R_dep} shows the difference of the free energy from the HOTRG.
By using the internal-line oversampling, the MDTRG exponentially converges to that from the HOTRG with no randomized method.
MDTRG requires the internal-line oversampling to obtain correct convergence without additional systematic error. 

\subsection{Approximation for the unit-cell tensor}

We also discuss the unit-cell tensor and approximated network region.
For the Triad-MDTRG, we define the triad representation $IJKLMNOP$ by the SVD which includes the whole unit-cell tensor $\Gamma$.
For the comparison, we also consider the Triad-MDTRG with the simple SVD of $E$ and $F$.
We can define $I = M = U^{(E)}\sqrt{s^{(E)}}$ and $J = N = V^{(E)}\sqrt{s^{(E)}}$ by the SVD of $E$ as $E = U^{(E)} s^{(E)}V^{(E)}$, and $L = P = U^{(F)}\sqrt{s^{(F)}}$ and $K = O = V^{(F)}\sqrt{s^{(F)}}$ by the SVD of $F$ as $F = U^{(F)} s^{(F)}V^{(F)}$.
Note that this simple SVD is not randomized.

\begin{figure}[tb!]
 \centering
\begin{minipage}[t]{0.5\columnwidth}
 \includegraphics[clip,width=1.0\columnwidth]{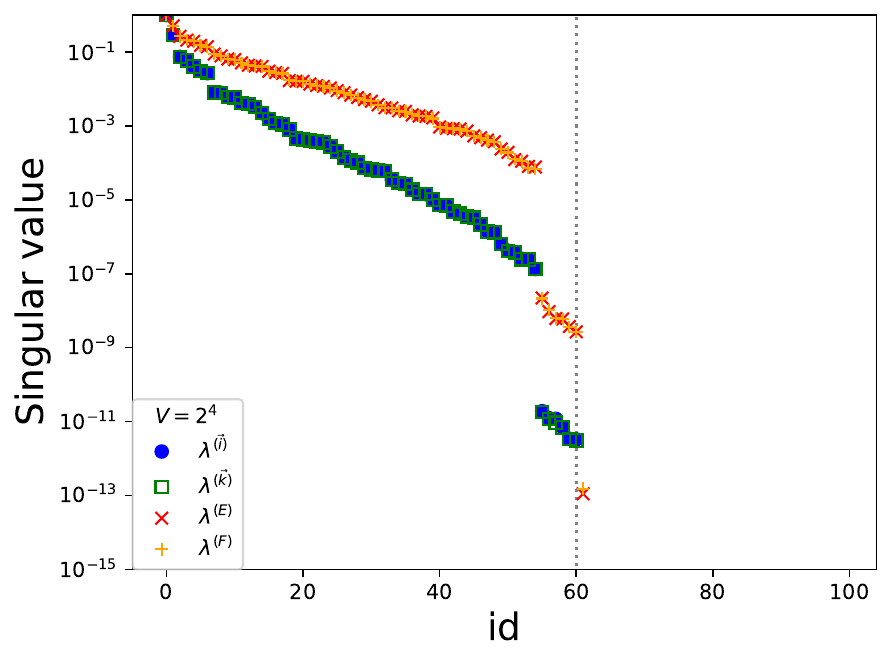}
\end{minipage}%
\begin{minipage}[t]{0.5\columnwidth}
 \includegraphics[clip,width=1.0\columnwidth]{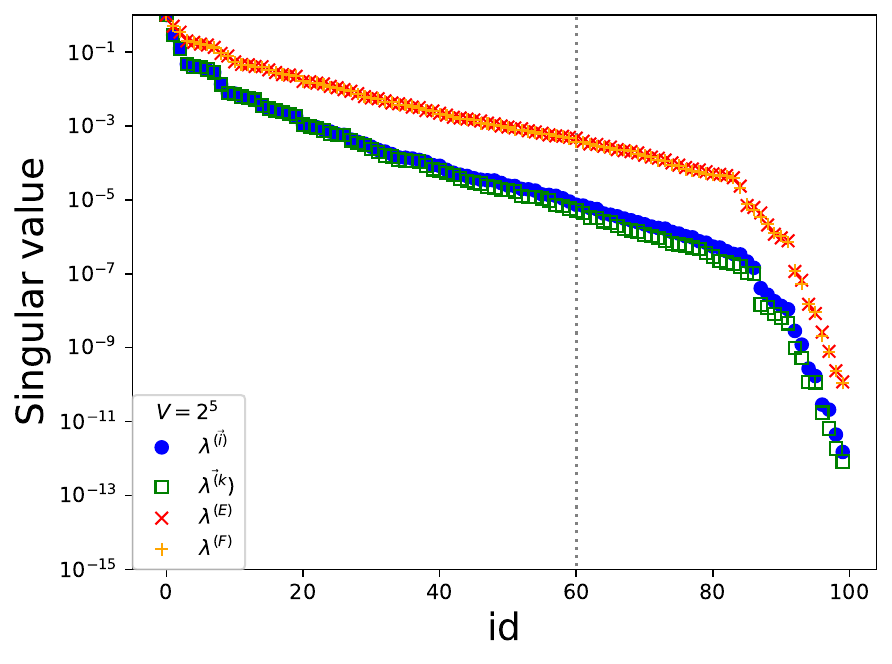}
\end{minipage}\\
\begin{minipage}[t]{0.5\columnwidth}
 \includegraphics[clip,width=1.0\columnwidth]{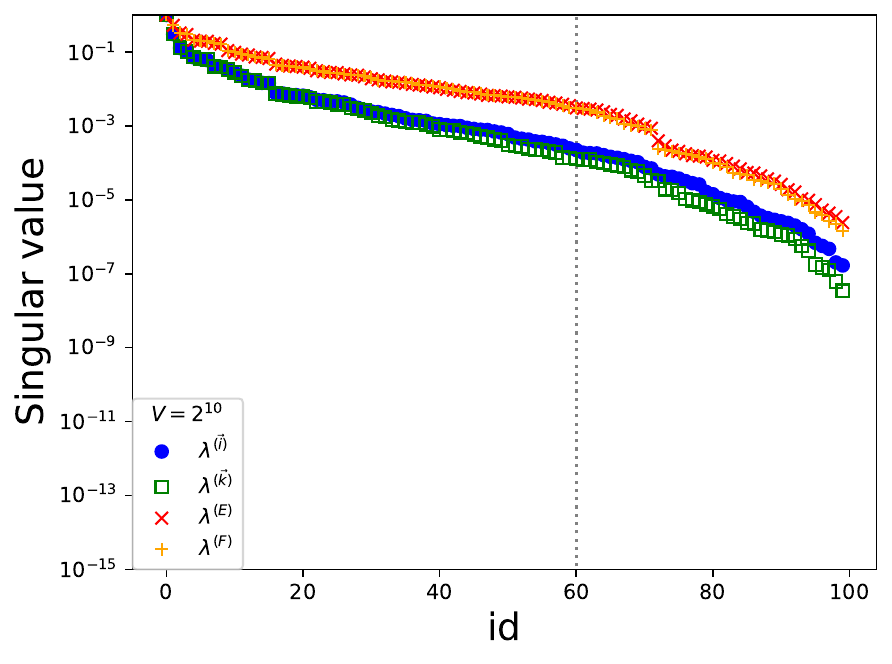}
\end{minipage}%
\begin{minipage}[t]{0.5\columnwidth}
 \includegraphics[clip,width=1.0\columnwidth]{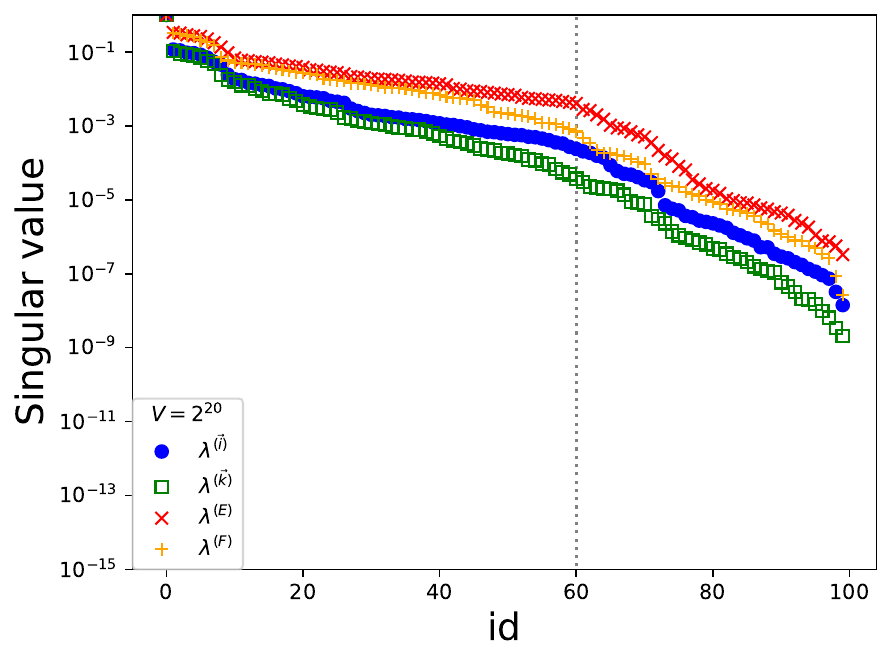}
\end{minipage}
\\
\begin{minipage}[t]{0.5\columnwidth}
 \includegraphics[clip,width=1.0\columnwidth]{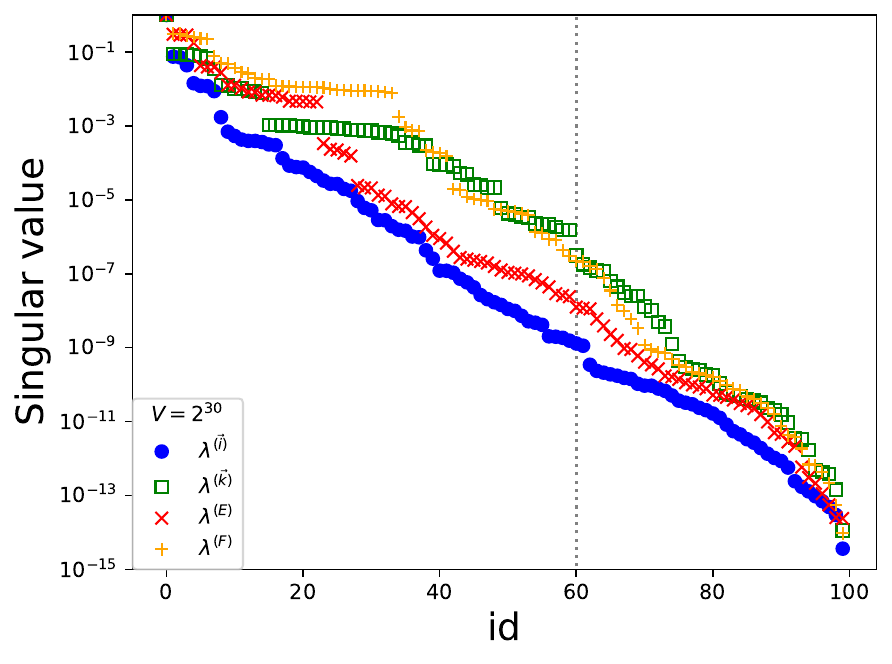}
\end{minipage}%
\begin{minipage}[t]{0.5\columnwidth}
 \includegraphics[clip,width=1.0\columnwidth]{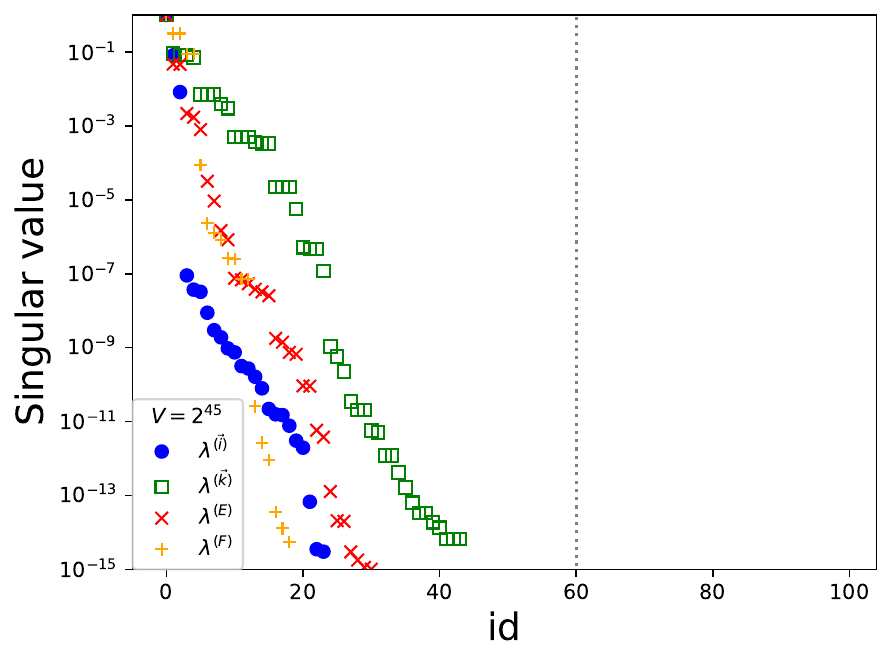}
\end{minipage}%
 \caption{The singular value to make the triad representation for the Triad-MDTRG $\lambda^{(\vec{i})}$ (circle) and $\lambda^{(\vec{k})}$ (square) for the isometry defined by the unit-cell tensor network $\Gamma^{(EFGH)}$, and the $\lambda^{(E)}$ (cross) and $\lambda^{(F)}$ (plus) for the isometry defined by the simple SVD of $E$ and $F$. The singular values are in descending order and normalized by the maximum singular value. The result in the volume $V = \{2^4, 2^5, 2^{10}, 2^{20}, 2^{30}, 2^{45}\}$ with the parameters $r=6$, $q=2$, and $D=10$ are plotted.
 The dotted line is $rD = 60$.
 }
 \label{SV_hierarcy}
\end{figure}

We define the $\lambda^{(E)} \equiv (s^{(E)})^2 $ and $\lambda^{(F)} \equiv (s^{(F)})^2$. Figure \ref{SV_hierarcy} shows the singular values of the Triad-MDTRG for $IJKLMNOP$ and the Triad-MDTRG with the simple SVD of the $E$ and $F$ in descending order with the parameter $r= 6$, $q=2$, and $D=10$.
The singular values of the Triad-MDTRG show faster decay than that of the Triad-MDTRG with the simple SVD, at least up to the step for the $V=2^{20}$.
In the $V=2^{45}$, the truncated singular values become zero.
The singular values of the Triad-MDTRG in $V=2^{30}$ are comparable to that of the Triad-MDTRG with the simple SVD.
We note that the truncation for the smaller volume is also essential for the larger volume 
because simulations at larger volumes contain contributions of truncations at each renormalization step.
Our result suggests that the isometry which comes from the unit-cell tensor network $\Gamma$ gives a better approximation than that from the simple SVD of the $E$ and $F$.

\subsection{Precision and computational cost}

\begin{figure}[tb!]
 \centering
 \includegraphics[clip,width=1.0\columnwidth]{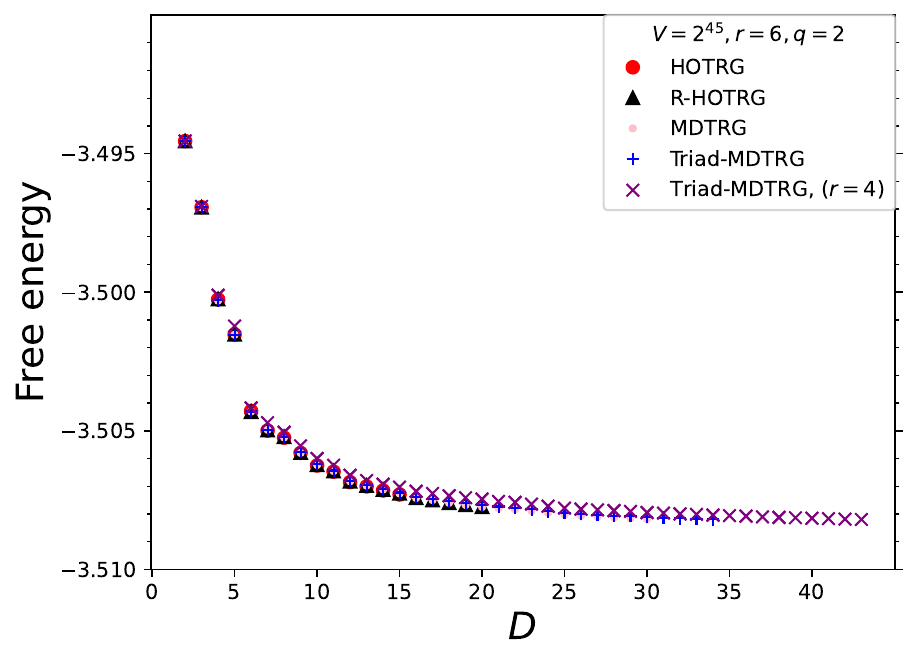}
 \caption{The free energy in the three-dimensional Ising model at a critical temperature depends on the truncated bond dimension $D$.}
 \label{3d_energy_D}
\end{figure}
\begin{figure}[tb!]
 \includegraphics[clip,width=1.0\columnwidth]{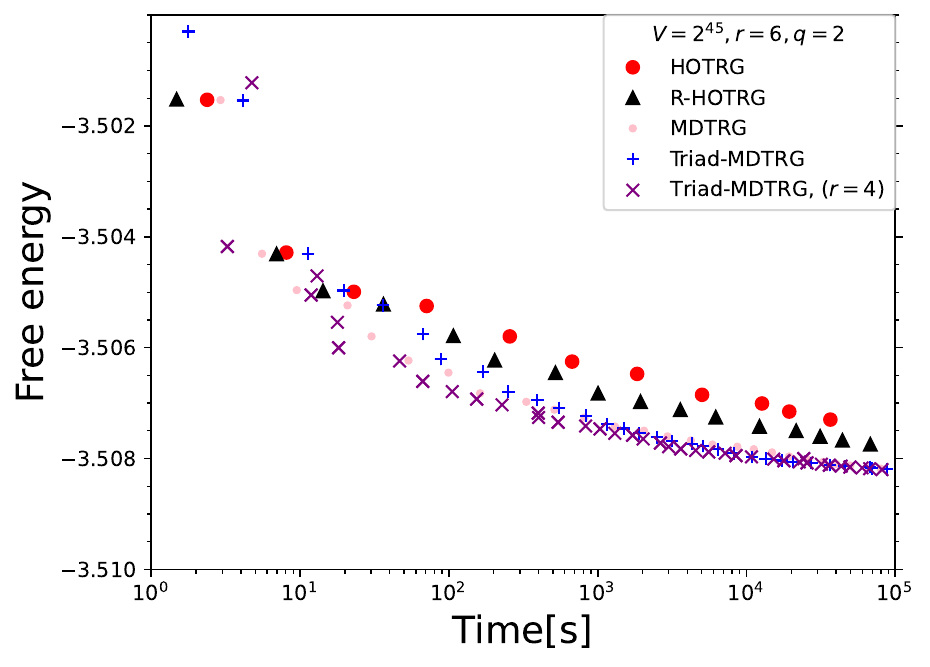}
 \caption{The free energy in the three-dimensional Ising model at a critical temperature depends on the processor time.}
 \label{3d_energy_Time}
\end{figure}

Figure \ref{3d_energy_D} shows the free energy depending on the truncated bond dimension $D$, and parameters $r = 6$, $q = 2$.
We also show the Triad-MDTRG with the parameter $r=4$ since the calculation time is reduced by factor $4^3/6^3\simeq 0.3$ although the systematic error from the R-SVD becomes large as shown in Fig. \ref{R_dep} for $D=10$.
The free energies from the R-HOTRG, MDTRG, and Triad-MDTRG converge to that from the HOTRG.
We also measure the total processor time and show the free energy depending on the processor time in Fig. \ref{3d_energy_Time}.
It clearly shows the cost reduction in the sufficiently large $D$ region.
Since the precision is in the same order as the HOTRG, the dominant part of the systematic error is only the truncation of the isometry.
This is an advantage of the R-HOTRG and MDTRG methods with the internal-line oversampling.

\begin{figure}[t]
 \centering
 \includegraphics[clip,width=1.0\columnwidth]{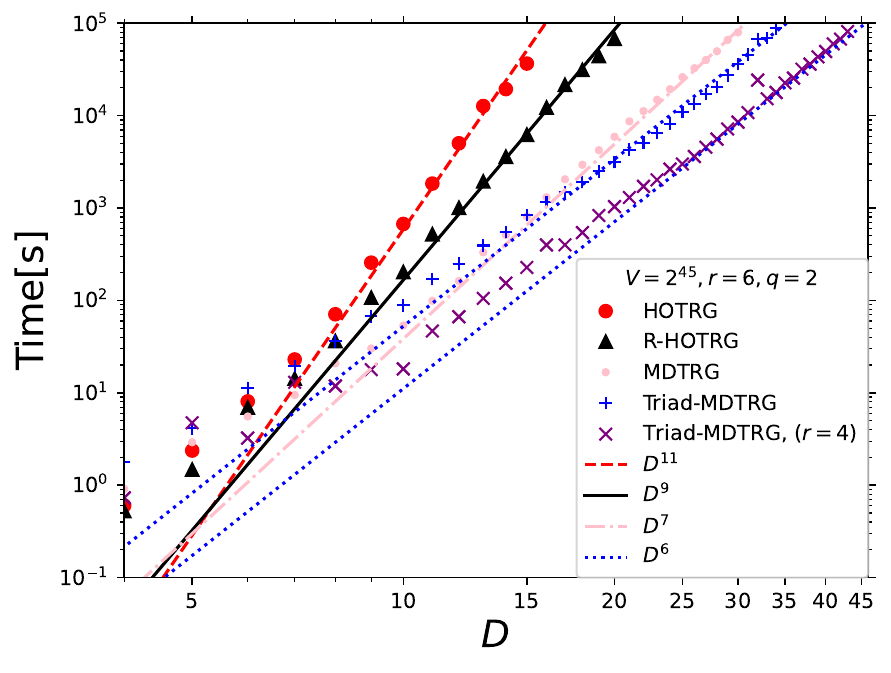}
 \caption{The processor time in three-dimensional Ising model at a critical temperature depends on the truncated bond dimension $D$.}
 \label{3d_scaling}
\end{figure}

This improvement may become more efficient in the large $D$ region because the scaling of $D$ is $O(D^{11})$ for HOTRG and $O(D^{9})$ for R-HOTRG, $O(D^7)$ for MDTRG, and $O(D^6)$ for Triad-MDTRG.
We demonstrate the scaling of the processor time with the $D$ in Fig. \ref{3d_scaling}.
The processor time grows as we expected.
It suggests that the R-HOTRG, MDTRG, and Triad-MDTRG produce almost the same precision result at the same truncated bond dimension $D$ with more and more reduced computational cost at larger $D$.
Our calculations demonstrate that the R-HOTRG and MDTRG could help us to obtain reliable and precise results at larger $D$, at least for the three-dimensional Ising model.

The results also imply that the systematic error of the R-HOTRG and MDTRG is originated from only the isometry step, which is the same as the HOTRG.
It helps us to obtain the physical quantities with no complicated systematic error which may come from the additional procedures to reduce the computational cost.
In this sense, the R-HOTRG is one of the simplest methods for the higher dimension to calculate the physical quantities.

\section{ Conclusion} \label{Sec:6}

In this work, we proposed the R-HOTRG method as a fundamental and most straightforward approach in higher dimensions with the randomized method.
The R-HOTRG helps us to perform the TRG method in higher dimensions without additional decompositions with $O(D^{3d})$ cost in $d$-dimension.
We also proposed the MDTRG and Triad-MDTRG to achieve $O(D^{2d+1})$ and $O(D^{d+3})$, respectively.
We introduce the internal-line oversampling and the isometry of the unit-cell tensor to improve the precision of the MDTRG and Triad-MDTRG.
These ideas help us understand the approximation in the TRG and can be applied to the general TRG method.
Our numerical calculation shows that the R-HOTRG, MDTRG, and Triad-MDTRG converge to the HOTRG at the same truncated bond dimension $D$, without any other systematic error.
Our study will be a fundamental knowledge of the TRG method in higher dimensions, both of the practical and theoretical work for the TRG formulation.

\section*{ACKNOWLEDGMENTS}
The author thank Shinji Takeda for fruitful knowledge of the loop-blocking method and Kei Suzuki for the detailed discussions.
The author also thanks Yasumichi Aoki and Yoshifumi Nakamura for encouraging this work.

\appendix
\section*{Appendix}
\renewcommand{\thesubsection}{\Alph{subsection}}

\subsection{Higher order tensor renormalization group} \label{App:1}
\renewcommand{\theequation}{A\arabic{equation}}
\setcounter{equation}{0}

We briefly introduce the original HOTRG \cite{HOSVD,HOTRG}. 
The HOTRG utilizes the idea of HOSVD to two-neighboring tensors.
In HOTRG, we try to find the coarse-grained tensor $A^\mathrm{(next)} _{XyX'y'}$ with the truncated indices $X$ and $X'$ from the two-neighboring tensor $\Gamma_{x_1x_2x' _1 x' _2 yy'} ^{(AA)}=\sum_{a=1} ^DA_{x_1ax' _1 y'}A_{x_2yx' _2a}$, where $D$ is the truncated bond dimension.
As the coarse-graining of $x$-direction, we have to find the isometry $U_{x_1x_2X}$ with the truncated bond dimension $D$ index $X$.
The isometry approximates the $D^2$ contraction of the $x_1$ and $x_2$ by index $X$ with the truncated bond dimension $D$.
In order to find the isometry for good approximation, we use the SVD.
We can define the isometry by using SVD as follows,
\begin{align}
&
\Gamma_{x_1x_2x' _1 x' _2 yy'} ^{(AA)}= \sum_{k=1} ^{D^2}U_{x_1x_2 k}s_kV_{x' _1 x' _2 yy' k},
\end{align}
where the unitary matrix $U$ and $V$ with the singuler value $s_k$.
We truncate the index $k = 1,...,D^2$ up to the truncated bond dimension $D$ ignoring the smaller singular values, and then the truncated tensor $U_{x_1x_2 k}$ is nothing but the isometry $U_{x_1x_2 X}$ by identifying the index $k$ as $X$.
Because the truncated singular values are smaller and the Frobenius norm is the sum of the singular values, this approximation is optimal considering the Frobenius norm of the $\Gamma$.

This direct SVD of $\Gamma$ requires $O(D^8)$ computational cost.
In order to reduce the cost, we apply the SVD to $\Gamma\Gamma^t$ instead of $\Gamma$.
\begin{align}
&
\sum_{x' _1, x' _2, y,y'=1}^D\Gamma_{[x_1x_2][x' _1 x' _2 yy']} ^{(AA)}\Gamma^{(AA)} _{[x_1 ^tx_2 ^t][x' _1 x' _2 yy']} \nonumber\\
&
=\sum_{k=1} ^{D^2}U_{x_1x_2 k}\lambda_kU _{x _1 ^tx _2 ^t k},
\end{align}
with the singular values $\lambda_k$.
The tensor $\Gamma\Gamma^t$ can be written by the tensor $A$ as follows.
\begin{align}
&
\Gamma_{[x_1x_2][x' _1 x' _2 yy']} ^{(AA)} \Gamma_{[x_1 ^tx_2 ^t][x' _1 x' _2 yy']} ^{(AA)} \nonumber\\
&
=\sum_{y,y',\tilde{y},\tilde{y}^t = 1} ^D A_{x_1\tilde{y}x' _1 y'}A_{x_2yx' _2\tilde{y}}A _{x_1 ^t\tilde{y}^tx' _1 {}^ty'} A _{x_2 ^tyx' _2 {}^t\tilde{y}^t}.
\end{align}
We also calculate the $x'$-direction in the same manner.
Since the system is homogeneous and periodic, we can choose the better one compared to the isometry of $x$ and $x'$ direction.
We choose isometry which has a smaller sum of the truncated singular values.
Finally we take the contraction of the tensors $A^{(\mathrm{next})} _{XyX'y'}=\sum_{x_1,x_2,x' _1, x' _2=1} ^D\Gamma_{x_1x_2x' _1 x' _2 yy'} ^{(AA)}U_{x_1x_2 X}U _{x' _1x' _2 X'}$ with the cost $O(D^7)$.
Note that the memory footprint of the contraction step is naively $O(D^5)$, but the loop-blocking technique reduces it to $O(D^4)$, corresponding to the tensor $A$.

The extension to the higher dimension is straightforward.
We consider the tensor $A_{xyzx'y'z'}$ and apply the same procedure to $x$ and $y$ direction.
The cost of HOTRG in the isometry step is $O(D^{2d+2})$ and the contraction step is $O(D^{4d -1})$.
The number of the tensor $A$ element is $D^{2d}$.

\subsection{Randomized singuler value decomposition} \label{App:2}
\renewcommand{\theequation}{B\arabic{equation}}
\setcounter{equation}{0}

We also briefly introduce the randomized SVD method \cite{rand_trunc,RandTRG}.
Let us consider the SVD of the $m\times n$ matrix $A_{ab} = \sum_{k = 1} ^{\mathrm{min}(m,n)}U_{ak}s_kV_{bk}$.
In the randomized SVD, we first prepare $n\times D$ random matrix $\Omega$ and define the $m\times D$ sample matrix $\Theta\equiv A\Omega$.
From the QR decomposition of the sample matrix $\Theta=QR$, we calculate the orthogonal matrix $Q$, and then the original matrix can be approximately written as $A \simeq Q(Q^\dagger A)$.
Finally, we take the SVD of $Q^\dagger A = \tilde{U}s_k V$, and then we get the unitary matrix $U \simeq Q\tilde{U}$.
We assume the index $D \ll m,n$.
The total cost is $O(mnD)$.
For $m=n=D^2$ case, simple SVD needs $O(D^6)$ cost, and the RSVD needs $O(D^5)$ cost.

Before the SVD of $Q^\dagger A$, the random sampling procedure is also powerful for the matrix-matrix product $BC$, where the $m\times l$ matrix $B$ and $l\times n$ matrix $C$ \cite{TriadTRG}.
The original matrix product needs the cost $O(mnl)$.
We assume the index $D \ll l,m,n$.
Substituting $A=BC$, we find the equation $BC \simeq Q((Q^\dagger B)C)$ with the cost $O((mn + ln + lm)D)$.
For $l=m=n=D^2$ case, this random sampling method needs a cost $O(D^5)$ while the original product needs $O(D^6)$.
In other words, randomized SVD is also helpful for the contraction step and the isometry step.

We note that the randomized SVD needs an oversampling parameter defined as the random matrix $\Omega$ set the $n\times rD$ with the coefficient $r$, to calculate reliable SVD up to the truncated bond dimension $D$.
More details of the randomized SVD are also discussed in \cite{RandTRG,TriadTRG}, and a theoretical discussion is shown in \cite{rand_trunc}.

\subsection{The isometry of the unit-cell tensor network} \label{App:3}
\renewcommand{\theequation}{C\arabic{equation}}
\setcounter{equation}{0}

We discuss the unit-cell tensor network and approximation by isometry and decomposition.
In order to discuss the unit-cell tensor network and approximated region, we introduce the Frobenius norm as a cost function that is minimized by the approximation by the isometry.
This cost function representation was introduced in the TNR method for the projective truncation, which approximates the sub-volume tensor network by smaller index \cite{TNR,LoopTNR,ProjectiveTRG}.
It is also discussed as the global optimization in \cite{GlobalTRG}.

Let us introduce this representation for the HOTRG with the unit-cell tensor $\Gamma^{(AA)}$.
In the HOTRG, we have to find the tensor $U$ to minimize the cost function defined by the Frobenius norm $|| \Gamma^{(AA)} - U^t\Gamma^{(AA)}U||$.
The HOTRG assumes the tensor $U$ comes from the SVD of the unit-cell tensor $\Gamma^{(AA)}$ as isometry.
Since the isometry satisfy $U^tU = 1$, the norm becomes
\begin{align}
&
|| \Gamma^{(AA)}- U^t\Gamma^{(AA)}U||
=
\sqrt{ 
\sum_{k = 1} ^{D^2} \lambda_k
- 
\sum_{k = 1} ^D \lambda_k } ,
\end{align}
We can choose the ordering of the singular values to minimize this norm.
If the singular values are ordered in descent ordering, this norm is minimized.

We should not confuse representation with projective truncation in the TNR.
The projective truncation does not need to assume that the tensor $U$ is the truncated unitary matrix from SVD of $\Gamma^{(AA)}$.
The TNR method finds the tensor $U$ by linearization and iterative calculation to minimize the cost function as the variational optimization. 
In this paper, we do not use the projective truncation method and assume the tensor $U$ is defined by the SVD of $\Gamma$, which is the same assumption as the HOTRG.

Another characteristic property of the HOTRG is that the unit-cell tensor $\Gamma$ is totally taken into account for each approximation step.
This property holds also in the R-HOTRG and MDTRG.
In these cases, we minimize the norm $||AA - U^tAAU||$ and $||EFGH-U^{(EFGH)t}EFGHU^{(EFGH)}||$ by the isometry $U$ for the unit-cell tensor network $\Gamma ^{(AA)}=AA$ and $\Gamma ^{(EFGH)} = EFGH$, respectively.
Since the R-HOTRG and MDTRG do not introduce the additional decompositions except the R-SVD for the contraction of the unit-cell tensor $\Gamma$ with isometry $U$, these approximations are the approximation of the whole unit-cell tensor network $\Gamma$.

We can apply this idea to the initial preparation of the tensor $E,F,G$, and $H$ in MDTRG.
Before we apply this idea, we consider the simple SVD for the tensor $A$, 
\begin{align}
&
A_{xyzx'y'z'} = \sum_{k = 1} ^{D^3}U^{(A)} _{x'y'z'k}s^{(A)} _kV^{(A)} _{xyzk}, 
\end{align}
we can represent this decomposition by the norm $|| A - U^{(A)t}AU^{(A)}||$ and define $E^{(A)} = U^{(A)}$, $F_{xyze} ^{(A)} = \sum_{x',y',z' = 1} ^DU^{(A)} _{x 'y 'z ' e}A_{xyzx'y'z'}$, $G_{x'y'z'g} ^{(A)}= \sum_{x,y,z = 1} ^DV^{(A)} _{x y z g}A_{xyzx'y'z'}$, and $H = V^{(A)}$.
This is also a good approximation if the truncated singular values $s^{(A)}$ are small enough.
On the other hand, this is not the approximation of the whole contribution from the unit-cell tensor network $\Gamma$, although the $A$ is part of the $\Gamma$.
Since the HOTRG only consider the approximation of the $\Gamma$, we would like to introduce the approximation with the same tensor network $\Gamma$.

In order to respect the contribution of the
$\Gamma_{x_1x_2x_1 ' x_2 'y_1y_2y_1 ' y_2 ' zz'} ^{(AA)}= \sum_{\tilde{z} = 1} ^DA_{x_1y_1 \tilde{z} x_1 ' y_1 ' z_1 '}A_{x_2y_2 z_2 x_2 ' y_2 ' \tilde{z}}$
for the initial preparation, we consider the SVD as followings,
\begin{align}
&
\sum_{x_1, x_2, x_2 ' , y_1, y_2, y_2 ' , z = 1} ^D
\Gamma_{x_1 x_2 x_1 ' x_2 ' y_1 y_2 y_1 ' y_2 ' z z'} ^{(AA)}\nonumber \\
&
\ \ \ \ \ \ \ \ \ \ \ \ \ \ \ \ \ \ \ \ \ \times
\Gamma_{x_1 x_2 x_1 ' {}^t x_2 ' y_1 y_2 y_1 ' {}^t y_2 ' z z' {}^t} ^{(AA)} \nonumber\\
&
\ \ \ \ \ \ \ \ \ \ \ \ \ \ \ \ \ \ \ \ \
= U^{(\vec{e})} _{x_1 ' y_1 ' z' k} \lambda^{(\vec{e})} _{k}U^{(\vec{e})} _{x_1 ' {}^ty_1 ' {}^tz' {}^t k},
\end{align}
and consider the cost function $||\Gamma^{(AA)} - U^{(e)t}\Gamma^{(AA)}U^{(e)}||$ for indices $x_1 ', y_1 '$, and $z_1 '$.
We define the $E = U^{(\vec{e})}$ and $F_{xyze} = \sum_{x',y',z' = 1} ^DU^{(\vec{e})} _{x 'y 'z ' e}A_{xyzx'y'z'}$.
In the same manner, we consider another SVD of $\Gamma^{(AA)}$ as $\Gamma^{(AA)}\Gamma^{(AA)t} = U^{(\vec{g})}\lambda^{(\vec{g})}U^{(\vec{g})t} $ with the cost function $||\Gamma ^{(AA)}- U^{(\vec{g})t}\Gamma^{(AA)}U^{(\vec{g})}||$ for indices $x_2, y_2, $ and $z_2$.
We define the $G_{x' y' z' g} = U^{(\vec{g})} _{x y z g}A_{xyzx'y'z'}$ and $H = U^{(\vec{g})}$.
Since this SVD comes from the unit-cell tensor $\Gamma$, this approximation takes into account the whole contribution of the $\Gamma$.
The order of the computational cost for the $U^{(\vec{e})}$ and $U^{(\vec{g})}$ is same as that of the $U^{(A)}$.

\bibliography{SVD}

\end{document}